%% file: main.tex
\documentclass[sigconf,screen]{acmart}

\AtBeginDocument{%
  }

\setcopyright{cc}
\setcctype{by-nc-nd}
\acmDOI{10.1145/3832783.3834544}
\acmYear{2026}
\copyrightyear{2026}
\acmISBN{979-8-4007-2882-2/2026/10}
\acmConference[ASE '26]{Proceedings of the 41st IEEE/ACM International Conference on Automated Software Engineering}{October 12--16, 2026}{Munich, Germany}
\acmBooktitle{Proceedings of the 41st IEEE/ACM International Conference on Automated Software Engineering (ASE '26), October 12--16, 2026, Munich, Germany}
\acmSubmissionID{ase26nier-p39-p}
\received{2026-05-12}
\received[accepted]{2026-07-02}

\input{utils/macros}

\begin{document}
\title{ECLAIR: A Causally-Grounded AI Framework for Scientific Discovery in Empirical Software Engineering}


\author{Alejandro Velasco}
\correspondingauthor
\orcid{0000-0002-4829-1017}
\affiliation{%
  \institution{William \& Mary}
  \city{Williamsburg}
  \country{USA}
}
\email{svelascodimate@wm.edu}

\author{Daniel Rodriguez-Cardenas}
\orcid{0000-0002-3238-1229}
\affiliation{%
  \institution{William \& Mary}
  \city{Williamsburg}
  \country{USA}
}
\email{dhrodriguezcar@wm.edu}

\author{Dipin Khati}
\orcid{0009-0008-4489-7733}
\affiliation{%
  \institution{William \& Mary}
  \city{Williamsburg}
  \country{USA}
}
\email{dkhati@wm.edu}

\author{David N. Palacio}
\orcid{0000-0001-6166-7595}
\affiliation{%
  \institution{Microsoft}
  \city{Redmond}
  \country{USA}
}
\email{davidnad@microsoft.com}

\author{Denys Poshyvanyk}
\orcid{0000-0002-5626-7586}
\affiliation{%
  \institution{William \& Mary}
  \city{Williamsburg}
  \country{USA}
}
\email{dposhyvanyk@wm.edu}

\renewcommand{\shortauthors}{Velasco et al.}

\input{text/00.Abstract}

\ccsdesc[500]{Computing methodologies}
\ccsdesc[500]{Software and its engineering}
\ccsdesc[500]{Computing methodologies~Artificial intelligence}
\ccsdesc[500]{Software and its engineering~Software creation and management}

\keywords{AI4SE, Research Methods, Empirical Research, \llms for Code}



\maketitle

\input{text/01.Introduction}

\input{text/03.Methodology}
\input{text/04.Case_study}

\input{text/05.Related_work}
\input{text/06.Conclusion}


\balance
\bibliographystyle{ACM-Reference-Format}
\bibliography{utils/references}


\end{document}

%% file: utils/macros.tex
\usepackage{tikz}
\usepackage{caption}
\usepackage{blindtext}
\usepackage{tcolorbox}
\usepackage[final]{pdfpages}
\usepackage{lipsum,multicol}
\usepackage{xcolor}
\usepackage{tikz}
\usepackage{listings}
\usepackage{enumitem}
\usepackage{hyperref}
\usepackage{amsfonts}
\usepackage{wrapfig}
\usepackage{subcaption} 
\usepackage{adjustbox}
\usepackage{colortbl}
\usepackage{fancybox}
\usepackage{multirow}
\usepackage[normalem]{ulem}
\useunder{\uline}{\ul}{}
\usepackage{enumitem}
\usepackage{amsmath}
\usepackage{booktabs} 
\usepackage{array}

\newcommand{\ie}{\textit{i.e.,}\xspace}
\newcommand{\eg}{\textit{e.g.,}\xspace}

\newcommand{\etal}{et al.\xspace}

\newcommand\revision[1]{{{#1}}}

\newtcolorbox{boxK}{
    fontupper = \small,
    sharpish corners, 
    boxrule = 0pt,
    toprule = 0pt, 
}

\newcommand{\figref}[1]{Fig.~\ref{#1}\xspace}
\newcommand{\tabref}[1]{Table~\ref{#1}\xspace}

\newcommand{\llm}{\textit{LLM}\xspace}
\newcommand{\llms}{\textit{LLMs}\xspace}

\newcommand{\sagent}{\textit{Scientific Agent}\xspace}

\definecolor{gradientplum}{RGB}{196,115,156}
\newcommand\prompt[1]{\texttt{\textcolor{gradientplum}{\textbf{#1}}}}

\newcommand{\framework}{\textit{ECLAIR}\xspace}

\definecolor{gradientplum}{RGB}{196,115,156}

\newcommand{\scm}{\textit{SCM}\xspace}

\newcommand{\ate}{\textit{ATE}\xspace}
\newcommand{\ates}{\textit{ATEs}\xspace}
\newcommand{\pearson}{$\rho$\xspace}

\newcommand{\approptoinn}[2]{\mathrel{\vcenter{
  \offinterlineskip\halign{\hfil$##$\cr
    #1\propto\cr\noalign{\kern2pt}#1\sim\cr\noalign{\kern-2pt}}}}}

\definecolor{gradientplum}{RGB}{196,115,156}
\newcommand\repository[1]{\textcolor{gradientplum}{\href{#1}{repository}}}

%% file: text/00.Abstract.tex
\begin{abstract}


The scientific method has long guided empirical research in Software Engineering (SE), but the complexity of modern software systems often hinders its systematic application. This paper introduces \framework, a causally grounded AI framework that integrates Large Language Models (\llms) into every stage of the scientific process, from hypothesis generation to analysis and interpretation. \framework treats \llms as active \textbf{scientific agents} operating under the principles of causal inference, within a human-in-the-loop design that safeguards against the risks of unsound automated reasoning. We demonstrate the framework through a case study examining how prompt design influences code generation accuracy in two \llms. Results show that, for both models, instruction-style, longer few-shot, and signature-augmented prompts yield small negative causal effects on accuracy, illustrating how causal reasoning provides a principled foundation for explaining \textit{why} software phenomena occur. This study presents the first causally grounded structured methodology for embedding \llms within the scientific method in SE, designed around the epistemological demands of empirical SE research, establishing a basis for rigorous AI-assisted research.

\end{abstract}

%% file: text/01.Introduction.tex
\section{Introduction}\label{sec:introduction}
\label{sec:introduction}

Francis Bacon (1620) argued that knowledge should be built through \textit{systematic observation} and \textit{reasoned experimentation}. Centuries later, the \textbf{scientific method} remains the foundation for transforming questions into understanding. At its core lies the search for \textbf{causal explanation}: to explain a phenomenon is to discover what causes it a goal formalized in Pearl's theory of causation through Structural Causal Models (SCMs) and the do-calculus~\cite{Pearl2009Causality}, which provide a mathematical basis for reasoning about interventions and their effects. Through this lens, scattered observations become structured knowledge, and empirical data are transformed into coherent reasoning. In Software Engineering (SE), the scientific method has guided decades of empirical research, shaping how researchers study the software life cycle, how teams collaborate, and how causal relationships are identified within complex systems.

Despite serving as a robust foundation for knowledge generation, the scientific method in SE is often constrained by the inherent complexity of the field. Modern software systems are \textit{large}, \textit{dynamic}, and \textit{collaborative}, making them difficult to observe, instrument, and analyze consistently. Empirical studies often depend on \textit{fragmented datasets}, \textit{inconsistent measurements}, and \textit{ad hoc tools} that hinder reproducibility. More fundamentally, SE research faces an epistemic challenge: \textit{connecting empirical evidence with the causal reasoning needed to explain it}. This challenge is well documented: most empirical SE studies establish statistical associations rather than causal relationships, and few employ the formal identification strategies required to support causal claims~\cite{docode, galeras, wohlin_experimentation_2012}. As the field grows increasingly data-intensive and interdisciplinary, it remains difficult to identify causal factors, formulate testable hypotheses, and link them to the mechanisms that govern software behavior. This gap limits our ability to draw \textit{credible causal conclusions} about which practices truly influence software engineering results.

Large Language Models (\llms) offer an opportunity to address this epistemic gap by supporting the reasoning processes that underpin empirical research. Trained in a vast corpus of code, documentation and scientific literature, \llms can assist not only in synthesizing prior evidence, but also in formulating hypotheses and structuring empirical analyzes. Across domains ranging from cancer biology~\cite{abdel-rehim_scientific_2025} (\eg proposing drug hypotheses later validated in laboratory experiments) and psychology~\cite{Tong_2024} to chemistry~\cite{liu_beyond_2025, boiko_autonomous_2023}, \llms have demonstrated the ability to generate and test causal hypotheses at a scale comparable to trained researchers~\cite{park_can_2024}, suggesting that under an appropriate methodological framework, they can act as \textbf{scientific agents} that extend human reasoning in the formulation and testing of causal claims.


\begin{figure*}[ht]
		\centering
  \vspace{-1.0em}
  \includegraphics[width=1\textwidth]{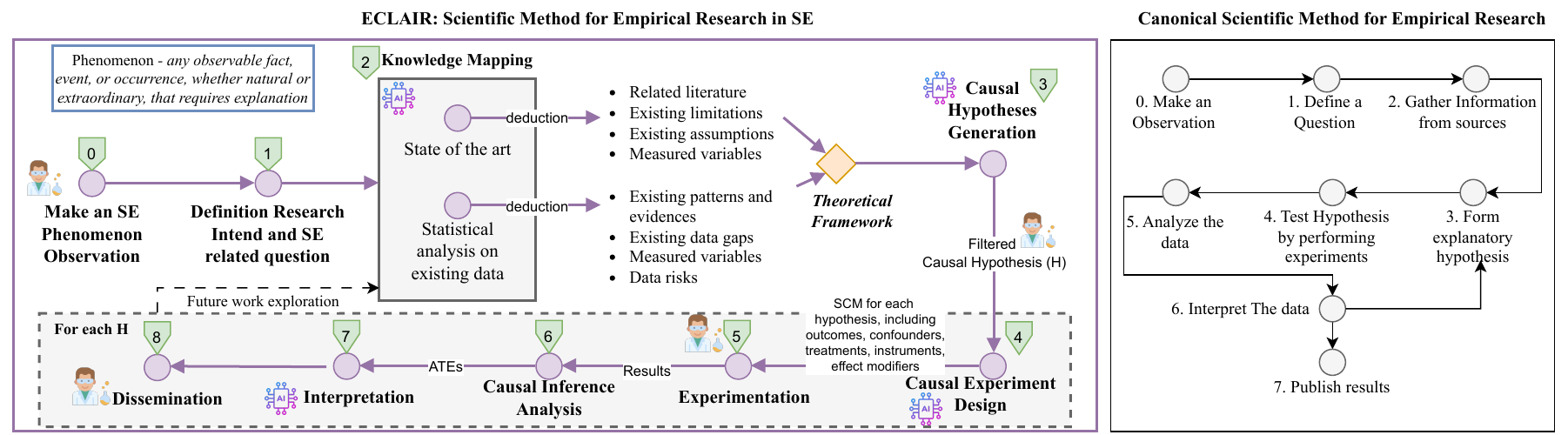}
  \vspace{-2.5em}
		\caption{Our \framework framework (left) contrasted with the canonical scientific method (right)}
    \label{fig:framework}
    \vspace{-1em}
\end{figure*}

This paper takes a step toward formalizing that potential within SE. We introduce \textbf{\framework} (\underline{E}mpirical-\underline{C}ausal \underline{L}LM-\underline{A}ugmented \underline{I}nference for SE \underline{R}esearch), a \textit{causally grounded AI framework} that integrates \llms into the scientific method, linking hypothesis generation with causal analysis through a transparent, human-in-the-loop process. This human-in-the-loop design is deliberate, as fully delegating scientific reasoning to automated agents risks perpetuating unsound research practices~\cite{jiang_badscientist_2025}, and \framework instead positions human judgment as a necessary checkpoint at each critical stage (\eg vetting hypotheses before experimentation, approving the causal model before effects are estimated). As illustrated in \figref{fig:framework}, \framework builds upon the canonical scientific method by embedding causal reasoning and \llm assistance across its stages, transforming it into a structured process for empirical discovery. Rather than treating \llms as passive instruments, \framework positions them as \textbf{scientific agents} whose contributions are \textit{documented}, \textit{constrained}, and \textit{evaluated} according to the principles of \textbf{causal inference}, enabling empirical research that is both scalable and scientifically rigorous.

To the best of our knowledge, \framework is the first framework to (i) operationalize Pearl's theory of causation as an executable, phase-by-phase scientific method for SE, rather than a domain-agnostic hypothesis generator, (ii) pair \llm assistance with mandatory human checkpoints per phase against unsound automated reasoning, and (iii) ground hypothesis and experiment design in SE-specific artifacts (\eg code, ASTs) instead of generic prompting. We demonstrate \framework through a case study that examines how prompt engineering influences model performance in code generation. All data, code and experimental configurations are publicly available in our replication package~\cite{repo}. Our contribution provides a foundation for using \llms not only to \textit{automate} research, but also to strengthen the reasoning process through which scientific discovery is achieved, grounded in the theory of causation.

%% file: text/03.Methodology.tex
\section{The \framework Framework For SE}
\label{sec:approach}

\framework is a human-in-the-loop framework that integrates \llms into the scientific method for SE, based on the theory of causation~\cite{Pearl2009Causality} and established methodological guidelines~\cite{armstrong_scientific_nodate}. \figref{fig:framework} contrasts \framework with the canonical scientific method across its \textbf{eight} sequential phases. For each phase below, we describe the researcher input given to the \sagent and its expected output; exact prompt templates and a detailed checklist are provided in our replication package~\cite{repo}.

\textbf{(1) Phenomenon Observation \& Research Intent.} The initial phase identifies SE phenomena requiring explanation, such as the \textit{influence of prompts on \llm code generation accuracy}. This phase does not involve the \sagent: the SE researcher selects a phenomenon, defines a research intention and question (optionally supported by a dataset), and applies expert judgment to prioritize problems of practical relevance and value~\cite{armstrong_scientific_nodate}.

\textbf{(2) Knowledge Mapping.} In this phase, the SE researcher selects a \llm (the \textbf{\sagent}) to construct the research knowledge base. \revision{ \textit{Input:} the agent uses the research input, dataset, and description or samples oriented by the defined research intention and available data (\eg types of code snippets).} The \sagent synthesizes the relevant literature, identifies limitations and assumptions of recent studies, and extracts patterns, variables, risks, and gaps from related datasets. \revision{ \textit{Output:} \sagent generates an observation deduction with the \textit{theoretical context} in a structured JSON file. The theoretical context captures the knowledge needed to formulate hypotheses about the SE phenomenon (\ie phenomenon scope, observable variables, quality risks, limitations)}. To ground synthesis in citable sources rather than parametric memory alone, curbing hallucination, this phase uses Retrieval-Augmented Generation (RAG) and agentic search for backward/forward snowballing over the literature.

\textbf{(3) Generation of Causal Hypotheses.} In this phase, observations become specific, testable causal hypotheses. \revision{\textit{Input:} \sagent uses phase (2) output and a prompt detailing hypothesis requirements (\ie measurable and testable variables, avoid speculative and abstract claims~\cite{repo})  }. All prompts are structurally grounded in Pearl's causal hierarchy~\cite{Pearl2009Causality}, requiring the \sagent to reason via \textit{why-based queries} (\eg ``why would $T$ change $O$?''), targeting the interventional/counterfactual rungs rather than associational \textit{what-is} questions, and to express each hypothesis in treatment/outcome/confounder form consistent with the \scm vocabulary used in Phase (4). This follows the ``\textit{Use Objective Designs}'' principle via Multiple Reasonable Hypotheses Testing (MRHT), considering established practices and competing alternatives~\cite{armstrong_scientific_nodate}. \revision{\textit{Output:} \sagent generates a set of hypotheses from the given phenomenon and following the input instructions and restrictions.} Given the risk of hallucination, each hypothesis is critically assessed by the SE researcher for soundness, validity, and feasibility before advancing to experimental design.

\textbf{(4) Causal Experiment Design.} This phase ensures the credibility of the study design. \revision{\textit{Input:} the SE researcher first selects the subset of hypotheses from Phase (3), then defines the relevant data constraints.} The \sagent generates a structural causal graph (\scm) specifying treatments, outcomes, and mediators. \revision{\textit{Output:} a JSON-encoded \scm specification per hypothesis.} The SE researcher then reviews and refines the proposed design, producing an experimental plan for each hypothesis. This reflects the scientific standard that experimental evidence, and meta-analyses thereof, provide the strongest validation~\cite{zhang_exploring_2025, baltes2025guidelinesempiricalstudiessoftware}.

\textbf{(5) Experimentation.} This phase executes the causal design to generate the empirical corpus for causal analysis. \revision{\textit{Input:} the \scm specification and the target dataset.} The SE researcher conducts controlled experiments per hypothesis, applying the defined treatments and measuring the outcomes, mediators, and confounders specified in the causal graph. The \sagent automates the experimentation workflow for consistency and reproducibility, \revision{producing logged model completions per treatment as \textit{output}}, while the SE researcher supervises and validates data completeness.

\textbf{(6) Causal Inference Analysis.} This phase applies Pearl's theory of causation~\cite{Pearl2009Causality,pearl_theoretical_2018,docode} to estimate causal effects from experimental data. \revision{\textit{Input:} the experimental corpus from Phase (5) and the \scm specification.} Within \framework, the \sagent estimates causal effects (\eg the Average Treatment Effect (\ate)) for each tested hypothesis while controlling for confounding bias, \revision{\textit{output:} \ate estimates and robustness/refutation results per method,} following the principles of valid, transparent, and logically consistent inference~\cite{armstrong_scientific_nodate}.

\textbf{(7) Interpretation.} In this phase, causal results become scientific insight. \revision{\textit{Input:} the \ate results from Phase (6) and related literature.} The \sagent synthesizes the findings and situates them within the existing literature, while the SE researcher critically evaluates their validity and relevance. \revision{\textit{Output:}} a consolidated interpretation that integrates validated findings and outlines open questions for future research.

\textbf{(8) Dissemination.} This final phase emphasizes Bacon's ``dissemination of useful findings,'' whereby conclusions must follow logically from evidence obtained through systematic observation. \revision{Taking Phase (7)'s interpretation as input, the} SE researcher reports results transparently, ensures reproducibility, and submits the work to peer review.

%% file: text/04.Case_study.tex
\section{A Study on Prompt Engineering}
\label{sec:case_study}


To demonstrate \framework, we conducted a case study on the phenomenon of how \textit{prompt design influences code generation accuracy}. In Phase (1), the SE researcher jointly fixed this phenomenon and a supporting dataset (detailed below), scoping the research question accordingly:

\vspace{-0.5em}
\begin{enumerate}[label=\textbf{RQ$_{\arabic*}$}, ref=\textbf{RQ$_{\arabic*}$}, wide, labelindent=5pt]\setlength{\itemsep}{0.2em}
      \item \label{rq:prompt_effect} \textit{What is the prompt design impact on the model's code generation accuracy?}
\end{enumerate}
\vspace{-0.5em}

\textbf{The \sagent.} We selected \textit{GPT-5} as the \sagent since it 
is to date the latest and largest \llm released by OpenAI. Larger models 
are preferred because their parameter capacity enhances reasoning depth, 
factual recall, and stability in complex analytical tasks~\cite{gpt5}.

\textbf{Dataset.} We use the \texttt{CodeText-Galeras}\cite{galeras} dataset, which pairs Python functions with natural language descriptions and structural metadata. The corpus aggregates samples from multiple open-source repositories and includes metrics such as token count, lines of code, \texttt{AST} depth, and documentation features. The combination of code and text enables the analysis of how prompt design affects code generation accuracy, as also explored in the dataset's original study~\cite{galeras}.

\textbf{Models.} We evaluate two instruction-tuned code generation models: \textit{Qwen/Qwen2.5-Coder-7B}~\cite{qwen} ($M_1$) and \textit{codellama/CodeLlama-7b-Python-hf}~\cite{codeLlama} ($M_2$). Both are optimized for program synthesis but differ in scope and specialization. $M_1$ is trained in multilingual code corpora with strong reasoning capabilities, while $M_2$ is a Python-focused variant of LLaMA. Both models share a comparable parameter count (7B), making their comparison controlled with respect to scale while contrasting in training strategy and domain specialization, which allows us to examine whether the causal effects of prompt design generalize across architectures or are model-specific. Like the dataset, $M_1$ and $M_2$ were fixed upfront rather than nominated during Experiment Design (\S Discussion).

\input{tables/results}

\textbf{Knowledge Mapping (Phase 2).} We began by contextualizing the phenomenon through a literature synthesis conducted with the \sagent. Recent studies \cite{fried2023incodergenerativemodelcode, madaan2023selfrefineiterativerefinementselffeedback, shinn2023reflexion, chen2023teachinglargelanguagemodels} show that instruction-style prompts and function signatures can influence metrics such as \texttt{pass@k} and \texttt{CodeBLEU}, although prompt effects often depended on model configuration and benchmark design, with limited causal validation. Within the same phase, the \sagent organized this evidence into a knowledge base by analyzing \texttt{CodeText-Galeras}, identifying measurable variables (\eg complexity, AST depth, token counts), confounders, and data gaps. Notably, it flagged that the dataset does not directly pair prompts with accuracy scores, a gap that Experiment Design (Phase 4) resolved by applying prompt treatments synthetically; the remaining variables were suitable for causal modeling.

Based on the mapped evidence, the \sagent formulated sixteen causal hypotheses linking prompt characteristics to accuracy metrics such as \textit{CodeBLEU}, \textit{Exact Match}, and \textit{Compilation Success} (see Appendix~\cite{repo}). Among them, $H_2$ (adding function signatures improves \texttt{CodeBLEU}) and $H_{14}$ (prompt truncation decreases \texttt{CodeBLEU}) had been explored previously~\cite{galeras}. For empirical evaluation, we selected three representative hypotheses: $H_{1}$ (instruction-style prompts improve \texttt{CodeBLEU}), $H_{11}$ (few-shot examples increase \texttt{CodeBLEU}), and $H_{13}$ (including function signatures benefits more complex functions).

For each hypothesis, the \sagent constructed a \scm defining treatments, outcomes, and confounders (see \tabref{tab:causal_results}). In $H_{1}$, the treatment was binary, with $T_0$ representing code-delimited prompts (control) and $T_1$ instruction-style prompts (intervention). In $H_{11}$, the treatment was continuous, representing the number of few-shot samples ($T_x$). In $H_{13}$, the treatment was binary, where $T_0$ excluded and $T_1$ included function signatures. For $H_{1}$ and $H_{11}$, structural and lexical features acted as confounders, while in $H_{13}$, complexity was modeled as an effect modifier.

Following the experimental designs, the SE researcher executed the experiments and estimated causal effects using the \textit{DoWhy} library~\cite{dowhy}. For each hypothesis ($H_{1}$, $H_{11}$, and $H_{13}$), the causal inference pipeline~\cite{docode} computed \ate and performed robustness checks using placebo, random common cause, data subset and unobserved confounder tests. The results, summarized in \tabref{tab:causal_results}, were consistent across both models and stable under \scm perturbations.

The estimated effects were then interpreted to assess how prompt structure influenced model performance. For $H_{1}$, the $\ate$ was negative ($-0.139$ for $M_1$, $-0.126$ for $M_2$), indicating that instruction-style prompts ($T_1$) reduced accuracy relative to code-delimited prompts ($T_0$). For $H_{11}$, the effect was also negative ($-0.001$ for $M_1$, $-0.028$ for $M_2$), suggesting that increasing few-shot demonstrations slightly decreased \texttt{CodeBLEU}. For $H_{13}$, including function signatures ($T_1$) produced a small negative effect ($-0.055$ for both models). In all cases, robustness tests confirmed that the effects were stable and not vulnerable to confounding bias.

\vspace{-0.5em}
\begin{boxK}
\vspace{-0.5em}
    \textit{\ref{rq:prompt_effect}}  
    For both $M_1$ and $M_2$, instruction-style, longer, and signature-augmented prompts yield small negative causal effects on accuracy.
    \vspace{-1.5em}
\end{boxK}
\vspace{-0.2em}

\textbf{Discussion.} Our results demonstrate the application of \framework in guiding the SE researcher from hypothesis generation to causal validation, ensuring that assumptions, treatments, and confounders are explicitly defined and systematically evaluated. The analysis revealed small but consistent negative effects across all hypotheses tested, showing that, within the scope of our study, instruction-style, longer, and signature-augmented prompts tend to reduce code generation accuracy for the two evaluated models. These findings challenge the expectation that adding structure or information to prompts improves model performance, instead suggesting that effectiveness depends on model architecture and training context. Unlike fixed causal pipelines, \framework constructs structural models dynamically from observed phenomena and evidence synthesized by the \sagent, improving transparency and validity. Designed to support researchers in answering causal questions that explain \textit{why} phenomena occur, \framework formalizes the reasoning required to identify and test causal mechanisms rather than describe correlations. As a simplification, the phenomenon, dataset, and candidate models were fixed jointly by the SE researcher in Phase (1), consistent with its definition (research intent \textit{may be supported by relevant datasets}), rather than having Knowledge Mapping propose datasets or Experiment Design nominate models conditioned on the generated hypotheses. The present case study intentionally addresses a focused causal hypothesis to illustrate the framework's end-to-end process; generalizability across more complex SE phenomena, larger datasets, and additional models remains an open direction, alongside a comparative evaluation between \sagent-generated outputs and those produced independently by SE researchers at each stage of the scientific process.

%% file: tables/results.tex
\begin{table*}[t]
\centering
\caption{Causal Effects (\ates) on CodeBLEU (\ie \ref{rq:prompt_effect})}
\label{tab:causal_results}
\vspace{-1.5em}
\setlength{\tabcolsep}{4pt}
\scalebox{0.85}{
\begin{tabular}{@{}lll l@{\hspace{2em}}rr@{\hspace{2em}}rr@{}}
\multicolumn{4}{c}{\textit{Causal Experimental Settings}} &
  \multicolumn{2}{c}{\textit{M1}~\cite{qwen}} &
  \multicolumn{2}{c}{\textit{M2}~\cite{codeLlama}} \\
\textbf{Id} & \textbf{Interv.} & \textbf{Treatments} & \textbf{Confounders} &
  \textbf{\pearson} & \textbf{\ate} & \textbf{\pearson} & \textbf{\ate} \\
\midrule
$H_{1}$ & \textit{binary} &
  \prompt{$T_0$: "\{incomplete\_method\}"\quad $T_1$: "\{instruction\} + \{incomplete\_method\}"} &
  $\mathcal{C}$ &
  \textbf{-0.575} & \cellcolor[HTML]{9673A6}{\color[HTML]{FFFFFF}-0.139} &
  \textbf{-0.515} & \cellcolor[HTML]{9673A6}{\color[HTML]{FFFFFF}-0.126} \\
\rowcolor[HTML]{EFEFEF}
$H_{11}$ & \textit{continuous} &
  \prompt{$T_x$: "\{instruction\} + \{x\_samples\} + \{incomplete\_method\}"} &
  $\mathcal{C}$ &
  \textbf{-0.163} & -0.001 &
  \textbf{-0.286} & -0.028 \\
$H_{13}$ & \textit{binary} &
  \prompt{$T_0$: "\{docstring\}"\quad $T_1$: "\{docstring\} + \{signature\}"} &
  $\mathcal{C} \setminus \{\texttt{complexity}\}$ &
  0.023 & -0.055 &
  -0.097 & -0.055 \\
\midrule
\multicolumn{8}{@{}p{1.05\textwidth}@{}}{\footnotesize
$\mathcal{C} = \{$\texttt{complexity}, \texttt{\#ast\_levels}, \texttt{\#ast\_nodes},
\texttt{\#ast\_errors}, \texttt{\#whitespaces}, \texttt{\#tokens}, \texttt{vocab\_size},
\texttt{\#words}, \texttt{\#identifiers}, \texttt{\#loc}, \texttt{prompt\_length},
\texttt{\#prompt\_tokens}$\}$.
The outcome is CodeBLEU for all hypotheses; \texttt{complexity} is modeled as an
effect modifier in $H_{13}$ and as a confounder otherwise.
\textbf{bold}: $-$ correlation;
\colorbox[HTML]{9673A6}{\color[HTML]{FFFFFF}purple}: $-$ causal effect.} \\
\end{tabular}
} 
\end{table*}

%% file: text/05.Related_work.tex
\section{Related Work}
\label{sec:related_work}

Recent systems such as Denario~\cite{villaescusanavarro2025denarioprojectdeepknowledge} and Kosmos~\cite{mitchener_kosmos_2025} demonstrate end-to-end scientific-discovery pipelines, from idea generation to paper drafting, across disciplines including astrophysics, biology, and chemistry~\cite{liu_beyond_2025,park_can_2024}. Google DeepMind's AlphaEvolve~\cite{novikov_alphaevolve_2025} combines evolutionary search with Gemini-powered \llms to autonomously discover novel algorithms. Wang \etal~\cite{wang_scientific_2023} survey AI's role in scientific discovery, Reddy \etal~\cite{reddy_scientific_2025} identify hypothesis generation as a key opportunity, and Zhou \etal~\cite{zhou_hypothesis_2024} show \llms can generate plausible hypotheses from literature, though without a formal causal structure or domain-specific methodology. None of these works addresses the epistemological demands of empirical SE research, where causal claims must be grounded in code-centric datasets and SE-specific confounders (\eg AST complexity, token structure). Within SE, Baltes \etal~\cite{baltes2025guidelinesempiricalstudiessoftware} propose methodological guidelines for \llm-involving studies, formalizing roles such as annotators, judges, and synthesis agents~\cite{ahmed2025llmsreplacemanualannotation, 11071936}.



Unlike fully automated systems that risk perpetuating unsound research practices~\cite{jiang_badscientist_2025}, \framework positions \llms as augmentative tools within a human-in-the-loop, causally grounded workflow that spans from hypothesis generation to validation. In contrast to prior work focused on task-specific assistance or domain-agnostic hypothesis generation, our approach integrates \llms into the epistemic structure of the scientific method itself, tailored to the specific challenges of empirical SE research.


%% file: text/06.Conclusion.tex
\section{Conclusions \& Future Plans}
\label{sec:conclusion}

This paper presented \framework, a causally grounded AI framework that integrates \llms into the scientific method for SE. By formalizing how \llms contributes to hypothesis generation, experimental design, and causal analysis, \framework establishes a structured and transparent foundation for empirical research. The case study on prompt engineering demonstrated how causal reasoning can be operationalized in practice, automating analytical steps while preserving human oversight and interpretability. In future work, we plan to extend \framework to additional SE domains, incorporate more advanced reasoning agents capable of critiquing causal assumptions, and expand the framework to support descriptive and temporal scientific questions beyond its current causal focus. Beyond SE, \framework aspires to contribute to the broader goal of AI-assisted science by providing tools that explain \textit{why}, \textit{how}, and \textit{when} complex software phenomena occur.


\section{Data Availability Statement}
\label{sec:conclusion}

All artifacts necessary to reproduce this study are publicly available in the repository at \cite{repo}, including the implementation of \framework, experimental scripts, prompts, and evaluation materials used in the case study. No proprietary or sensitive data were used.

%% file: utils/references.bib
@misc{ahmed2025llmsreplacemanualannotation,
      title={Can LLMs Replace Manual Annotation of Software Engineering Artifacts?}, 
      author={Toufique Ahmed and Premkumar Devanbu and Christoph Treude and Michael Pradel},
      year={2025},
      eprint={2408.05534},
      archivePrefix={arXiv},
      primaryClass={cs.SE},
      url={https://arxiv.org/abs/2408.05534}, 
}

@misc{baltes2025guidelinesempiricalstudiessoftware,
      title={Guidelines for Empirical Studies in Software Engineering involving Large Language Models}, 
      author={Sebastian Baltes and Florian Angermeir and Chetan Arora and Marvin Muñoz Barón and Chunyang Chen and Lukas Böhme and Fabio Calefato and Neil Ernst and Davide Falessi and Brian Fitzgerald and Davide Fucci and Marcos Kalinowski and Stefano Lambiase and Daniel Russo and Mircea Lungu and Lutz Prechelt and Paul Ralph and Rijnard van Tonder and Christoph Treude and Stefan Wagner},
      year={2025},
      eprint={2508.15503},
      archivePrefix={arXiv},
      primaryClass={cs.SE},
      url={https://arxiv.org/abs/2508.15503}, 
}

@article{abdel-rehim_scientific_2025,
	title = {Scientific hypothesis generation by large language models: laboratory validation in breast cancer treatment},
	volume = {22},
	shorttitle = {Scientific hypothesis generation by large language models},
	url = {https://royalsocietypublishing.org/doi/10.1098/rsif.2024.0674},
	doi = {10.1098/rsif.2024.0674},
	number = {227},
	urldate = {2025-10-28},
	journal = {Journal of The Royal Society Interface},
	author = {Abdel-Rehim, Abbi and Zenil, Hector and Orhobor, Oghenejokpeme and Fisher, Marie and Collins, Ross J. and Bourne, Elizabeth and Fearnley, Gareth W. and Tate, Emma and Smith, Holly X. and Soldatova, Larisa N. and King, Ross},
	month = jun,
	year = {2025},
	note = {Publisher: Royal Society},
	pages = {20240674},
}

@article{Tong_2024,
   title={Automating psychological hypothesis generation with AI: when large language models meet causal graph},
   volume={11},
   ISSN={2662-9992},
   url={http://dx.doi.org/10.1057/s41599-024-03407-5},
   DOI={10.1057/s41599-024-03407-5},
   number={1},
   journal={Humanities and Social Sciences Communications},
   publisher={Springer Science and Business Media LLC},
   author={Tong, Song and Mao, Kai and Huang, Zhen and Zhao, Yukun and Peng, Kaiping},
   year={2024},
   month=jul }

@article{boiko_autonomous_2023,
	title = {Autonomous chemical research with large language models},
	volume = {624},
	copyright = {2023 The Author(s)},
	issn = {1476-4687},
	url = {https://www.nature.com/articles/s41586-023-06792-0},
	doi = {10.1038/s41586-023-06792-0},
	language = {en},
	number = {7992},
	urldate = {2025-10-28},
	journal = {Nature},
	author = {Boiko, Daniil A. and MacKnight, Robert and Kline, Ben and Gomes, Gabe},
	month = dec,
	year = {2023},
	note = {Publisher: Nature Publishing Group},
	pages = {570--578},
}

@article{liu_beyond_2025,
	title = {Beyond designer's knowledge: {Generating} materials design hypotheses via a large language model},
	volume = {297},
	issn = {1359-6454},
	shorttitle = {Beyond designer's knowledge},
	url = {https://www.sciencedirect.com/science/article/pii/S1359645425005932},
	doi = {10.1016/j.actamat.2025.121307},
	urldate = {2025-10-28},
	journal = {Acta Materialia},
	author = {Liu, Quanliang and Polak, Maciej P. and Kim, So Yeon and Shuvo, MD Al Amin and Deodhar, Hrishikesh Shridhar and Han, Jeongsoo and Morgan, Dane and Oh, Hyunseok},
	month = sep,
	year = {2025},
	pages = {121307},
}

@article{park_can_2024,
	title = {Can {ChatGPT} be used to generate scientific hypotheses?},
	volume = {10},
	issn = {2352-8478},
	url = {https://www.sciencedirect.com/science/article/pii/S2352847823001557},
	doi = {10.1016/j.jmat.2023.08.007},
	number = {3},
	urldate = {2025-10-28},
	journal = {Journal of Materiomics},
	author = {Park, Yang Jeong and Kaplan, Daniel and Ren, Zhichu and Hsu, Chia-Wei and Li, Changhao and Xu, Haowei and Li, Sipei and Li, Ju},
	month = may,
	year = {2024},
	pages = {578--584},
}

@misc{fried2023incodergenerativemodelcode,
      title={InCoder: A Generative Model for Code Infilling and Synthesis}, 
      author={Daniel Fried and Armen Aghajanyan and Jessy Lin and Sida Wang and Eric Wallace and Freda Shi and Ruiqi Zhong and Wen-tau Yih and Luke Zettlemoyer and Mike Lewis},
      year={2023},
      eprint={2204.05999},
      archivePrefix={arXiv},
      primaryClass={cs.SE},
      url={https://arxiv.org/abs/2204.05999}, 
}

@misc{madaan2023selfrefineiterativerefinementselffeedback,
      title={Self-Refine: Iterative Refinement with Self-Feedback}, 
      author={Aman Madaan and Niket Tandon and Prakhar Gupta and Skyler Hallinan and Luyu Gao and Sarah Wiegreffe and Uri Alon and Nouha Dziri and Shrimai Prabhumoye and Yiming Yang and Shashank Gupta and Bodhisattwa Prasad Majumder and Katherine Hermann and Sean Welleck and Amir Yazdanbakhsh and Peter Clark},
      year={2023},
      eprint={2303.17651},
      archivePrefix={arXiv},
      primaryClass={cs.CL},
      url={https://arxiv.org/abs/2303.17651}, 
}

@inproceedings{shinn2023reflexion,
author = {Shinn, Noah and Cassano, Federico and Gopinath, Ashwin and Narasimhan, Karthik and Yao, Shunyu},
title = {Reflexion: language agents with verbal reinforcement learning},
year = {2023},
publisher = {Curran Associates Inc.},
address = {Red Hook, NY, USA},
doi = {10.52202/075280-0377},
booktitle = {Proceedings of the 37th International Conference on Neural Information Processing Systems},
articleno = {377},
numpages = {19},
location = {New Orleans, LA, USA},
series = {NIPS '23}
}

@misc{chen2023teachinglargelanguagemodels,
      title={Teaching Large Language Models to Self-Debug}, 
      author={Xinyun Chen and Maxwell Lin and Nathanael Schärli and Denny Zhou},
      year={2023},
      eprint={2304.05128},
      archivePrefix={arXiv},
      primaryClass={cs.CL},
      url={https://arxiv.org/abs/2304.05128}, 
}

@misc{galeras,
      title={Benchmarking Causal Study to Interpret Large Language Models for Source Code}, 
      author={Daniel Rodriguez-Cardenas and David N. Palacio and Dipin Khati and Henry Burke and Denys Poshyvanyk},
      year={2023},
      eprint={2308.12415},
      archivePrefix={arXiv},
      primaryClass={cs.SE},
      url={https://arxiv.org/abs/2308.12415}, 
}

@article{docode,
   title={Toward a Theory of Causation for Interpreting Neural Code Models},
   volume={50},
   ISSN={2326-3881},
   url={http://dx.doi.org/10.1109/TSE.2024.3379943},
   DOI={10.1109/tse.2024.3379943},
   number={5},
   journal={IEEE Transactions on Software Engineering},
   publisher={Institute of Electrical and Electronics Engineers (IEEE)},
   author={Nader Palacio, David and Velasco, Alejandro and Cooper, Nathan and Rodriguez, Alvaro and Moran, Kevin and Poshyvanyk, Denys},
   year={2024},
   month=may, pages={1215–1243} }

@ARTICLE{11071936,
  author={Crupi, Giuseppe and Tufano, Rosalia and Velasco, Alejandro and Mastropaolo, Antonio and Poshyvanyk, Denys and Bavota, Gabriele},
  journal={IEEE Transactions on Software Engineering}, 
  title={On the Effectiveness of LLM-as-a-Judge for Code Generation and Summarization}, 
  year={2025},
  volume={51},
  number={8},
  pages={2329-2345},
  doi={10.1109/TSE.2025.3586082}}

@software{repo,
  author  = {{SEMERU Lab}},
  title   = {{ci4sesci: Causal Inference for Software Science}},
  year    = {2025},
  url     = {https://github.com/WM-SEMERU/ci4sesci},
  note    = {GitHub repository}
}

@book{Pearl2009Causality,
    title = {{Causality: models, reasoning, and inference}},
    year = {2009},
    author = {Pearl, Judea},
    isbn = {978-0-521-89560-0},
    doi = {10.1017/CBO9780511803161}
}

@misc{gpt5,
	title = {Introducing {GPT}-5},
	url = {https://openai.com/index/introducing-gpt-5/},
	language = {en-US},
	urldate = {2025-11-04},
	month = nov,
	year = {2025},
}

@book{armstrong_scientific_nodate,
  title     = {The Scientific Method: A Guide to Finding Useful Knowledge},
  author    = {Armstrong, J. Scott and Green, Kesten C.},
  year      = {2022},
  publisher = {Cambridge University Press},
  isbn      = {978-1-316-51516-7},
  doi       = {10.1017/9781009092265},
}

@article{zhang_exploring_2025,
    title = {Exploring the role of large language models in the scientific method: from hypothesis to discovery},
    volume = {1},
    issn = {3005-1460},
    shorttitle = {Exploring the role of large language models in the scientific method},
    url = {https://www.nature.com/articles/s44387-025-00019-5},
    doi = {10.1038/s44387-025-00019-5},
    language = {en},
    number = {1},
    urldate = {2025-11-05},
    journal = {npj Artificial Intelligence},
    author = {Zhang, Yanbo and Khan, Sumeer A. and Mahmud, Adnan and Yang, Huck and Lavin, Alexander and Levin, Michael and Frey, Jeremy and Dunnmon, Jared and Evans, James and Bundy, Alan and Dzeroski, Saso and Tegner, Jesper and Zenil, Hector},
    month = aug,
    year = {2025},
    pages = {14},
}

@misc{qwen,
      title={Qwen Technical Report}, 
      author={Jinze Bai and Shuai Bai and Yunfei Chu and Zeyu Cui and Kai Dang and Xiaodong Deng and Yang Fan and Wenbin Ge and Yu Han and Fei Huang and Binyuan Hui and Luo Ji and Mei Li and Junyang Lin and Runji Lin and Dayiheng Liu and Gao Liu and Chengqiang Lu and Keming Lu and Jianxin Ma and Rui Men and Xingzhang Ren and Xuancheng Ren and Chuanqi Tan and Sinan Tan and Jianhong Tu and Peng Wang and Shijie Wang and Wei Wang and Shengguang Wu and Benfeng Xu and Jin Xu and An Yang and Hao Yang and Jian Yang and Shusheng Yang and Yang Yao and Bowen Yu and Hongyi Yuan and Zheng Yuan and Jianwei Zhang and Xingxuan Zhang and Yichang Zhang and Zhenru Zhang and Chang Zhou and Jingren Zhou and Xiaohuan Zhou and Tianhang Zhu},
      year={2023},
      eprint={2309.16609},
      archivePrefix={arXiv},
      primaryClass={cs.CL},
      url={https://arxiv.org/abs/2309.16609}, 
}

@misc{codeLlama,
      title={Code Llama: Open Foundation Models for Code}, 
      author={Baptiste Rozière and Jonas Gehring and Fabian Gloeckle and Sten Sootla and Itai Gat and Xiaoqing Ellen Tan and Yossi Adi and Jingyu Liu and Romain Sauvestre and Tal Remez and Jérémy Rapin and Artyom Kozhevnikov and Ivan Evtimov and Joanna Bitton and Manish Bhatt and Cristian Canton Ferrer and Aaron Grattafiori and Wenhan Xiong and Alexandre Défossez and Jade Copet and Faisal Azhar and Hugo Touvron and Louis Martin and Nicolas Usunier and Thomas Scialom and Gabriel Synnaeve},
      year={2024},
      eprint={2308.12950},
      archivePrefix={arXiv},
      primaryClass={cs.CL},
      url={https://arxiv.org/abs/2308.12950}, 
}

@misc{pearl_theoretical_2018,
    title = {Theoretical {Impediments} to {Machine} {Learning} {With} {Seven} {Sparks} from the {Causal} {Revolution}},
    url = {http://arxiv.org/abs/1801.04016},
    doi = {10.48550/arXiv.1801.04016},
    language = {en},
    urldate = {2025-02-04},
    publisher = {arXiv},
    author = {Pearl, Judea},
    month = jan,
    year = {2018},
    note = {arXiv:1801.04016 [cs]},
}

@misc{villaescusanavarro2025denarioprojectdeepknowledge,
      title={The Denario project: Deep knowledge AI agents for scientific discovery}, 
      author={Francisco Villaescusa-Navarro and Boris Bolliet and Pablo Villanueva-Domingo and Adrian E. Bayer and Aidan Acquah and Chetana Amancharla and Almog Barzilay-Siegal and Pablo Bermejo and Camille Bilodeau and Pablo Cárdenas Ramírez and Miles Cranmer and Urbano L. França and ChangHoon Hahn and Yan-Fei Jiang and Raul Jimenez and Jun-Young Lee and Antonio Lerario and Osman Mamun and Thomas Meier and Anupam A. Ojha and Pavlos Protopapas and Shimanto Roy and David N. Spergel and Pedro Tarancón-Álvarez and Ujjwal Tiwari and Matteo Viel and Digvijay Wadekar and Chi Wang and Bonny Y. Wang and Licong Xu and Yossi Yovel and Shuwen Yue and Wen-Han Zhou and Qiyao Zhu and Jiajun Zou and Íñigo Zubeldia},
      year={2025},
      eprint={2510.26887},
      archivePrefix={arXiv},
      primaryClass={cs.AI},
      url={https://arxiv.org/abs/2510.26887}, 
}

@misc{novikov_alphaevolve_2025,
    title = {{AlphaEvolve}: {A} coding agent for scientific and algorithmic discovery},
    shorttitle = {{AlphaEvolve}},
    url = {http://arxiv.org/abs/2506.13131},
    doi = {10.48550/arXiv.2506.13131},
    language = {en},
    urldate = {2025-11-05},
    publisher = {arXiv},
    author = {Novikov, Alexander and Vũ, Ngân and Eisenberger, Marvin and Dupont, Emilien and Huang, Po-Sen and Wagner, Adam Zsolt and Shirobokov, Sergey and Kozlovskii, Borislav and Ruiz, Francisco J. R. and Mehrabian, Abbas and Kumar, M. Pawan and See, Abigail and Chaudhuri, Swarat and Holland, George and Davies, Alex and Nowozin, Sebastian and Kohli, Pushmeet and Balog, Matej},
    month = jun,
    year = {2025},
    note = {arXiv:2506.13131 [cs]},
}

@misc{jiang_badscientist_2025,
    title = {{BadScientist}: {Can} a {Research} {Agent} {Write} {Convincing} but {Unsound} {Papers} that {Fool} {LLM} {Reviewers}?},
    shorttitle = {{BadScientist}},
    url = {http://arxiv.org/abs/2510.18003},
    doi = {10.48550/arXiv.2510.18003},
    language = {en},
    urldate = {2025-10-30},
    publisher = {arXiv},
    author = {Jiang, Fengqing and Feng, Yichen and Li, Yuetai and Niu, Luyao and Alomair, Basel and Poovendran, Radha},
    month = oct,
    year = {2025},
    note = {arXiv:2510.18003 [cs]},
}

@misc{mitchener_kosmos_2025,
    title = {Kosmos: {An} {AI} {Scientist} for {Autonomous} {Discovery}},
    shorttitle = {Kosmos},
    url = {http://arxiv.org/abs/2511.02824},
    doi = {10.48550/arXiv.2511.02824},
    language = {en},
    urldate = {2025-11-05},
    publisher = {arXiv},
    author = {Mitchener, Ludovico and Yiu, Angela and Chang, Benjamin and Bourdenx, Mathieu and Nadolski, Tyler and Sulovari, Arvis and Landsness, Eric C. and Barabasi, Daniel L. and Narayanan, Siddharth and Evans, Nicky and Reddy, Shriya and Foiani, Martha and Kamal, Aizad and Shriver, Leah P. and Cao, Fang and Wassie, Asmamaw T. and Laurent, Jon M. and Melville-Green, Edwin and Caldas, Mayk and Bou, Albert and Roberts, Kaleigh F. and Zagorac, Sladjana and Orr, Timothy C. and Orr, Miranda E. and Zwezdaryk, Kevin J. and Ghareeb, Ali E. and McCoy, Laurie and Gomes, Bruna and Ashley, Euan A. and Duff, Karen E. and Buonassisi, Tonio and Rainforth, Tom and Bateman, Randall J. and Skarlinski, Michael and Rodriques, Samuel G. and Hinks, Michaela M. and White, Andrew D.},
    month = nov,
    year = {2025},
    note = {arXiv:2511.02824 [cs]},
}

@article{dowhy,
  title   = {DoWhy: An End-to-End Library for Causal Inference},
  author  = {Sharma, Amit and Kiciman, Emre},
  journal = {arXiv preprint arXiv:2011.04216},
  year    = {2020},
  doi     = {10.48550/arXiv.2011.04216}
}

@book{wohlin_experimentation_2012,
    author = {Wohlin, Claes and Runeson, Per and H{\"o}st, Martin and Ohlsson, Magnus C. and Regnell, Bj{\"r}n and Wessl{\'e}n, Anders},
    title = {Experimentation in Software Engineering},
    year = {2012},
    isbn = {978-3-642-29044-2},
    publisher = {Springer},
    doi = {10.1007/978-3-642-29044-2}
}

@inproceedings{reddy_scientific_2025,
author = {Reddy, Chandan K. and Shojaee, Parshin},
title = {Towards scientific discovery with generative AI: progress, opportunities, and challenges},
year = {2025},
isbn = {978-1-57735-897-8},
publisher = {AAAI Press},
url = {https://doi.org/10.1609/aaai.v39i27.35084},
doi = {10.1609/aaai.v39i27.35084},
booktitle = {Proceedings of the Thirty-Ninth AAAI Conference on Artificial Intelligence and Thirty-Seventh Conference on Innovative Applications of Artificial Intelligence and Fifteenth Symposium on Educational Advances in Artificial Intelligence},
articleno = {3186},
numpages = {9},
series = {AAAI'25/IAAI'25/EAAI'25}
}

@ARTICLE{wang_scientific_2023,
       author = {{Wang}, Hanchen and {Fu}, Tianfan and {Du}, Yuanqi and {Gao}, Wenhao and {Huang}, Kexin and {Liu}, Ziming and {Chandak}, Payal and {Liu}, Shengchao and {Van Katwyk}, Peter and {Deac}, Andreea and {Anandkumar}, Anima and {Bergen}, Karianne and {Gomes}, Carla P. and {Ho}, Shirley and {Kohli}, Pushmeet and {Lasenby}, Joan and {Leskovec}, Jure and {Liu}, Tie-Yan and {Manrai}, Arjun and {Marks}, Debora and {Ramsundar}, Bharath and {Song}, Le and {Sun}, Jimeng and {Tang}, Jian and {Veli{\v{c}}kovi{\'c}}, Petar and {Welling}, Max and {Zhang}, Linfeng and {Coley}, Connor W. and {Bengio}, Yoshua and {Zitnik}, Marinka},
        title = "{Scientific discovery in the age of artificial intelligence}",
      journal = {\nat},
         year = 2023,
        month = aug,
       volume = {620},
       number = {7972},
        pages = {47-60},
          doi = {10.1038/s41586-023-06221-2},
       adsurl = {https://ui.adsabs.harvard.edu/abs/2023Natur.620...47W}
}

@inproceedings{zhou_hypothesis_2024,
    title = "Hypothesis Generation with Large Language Models",
    author = "Zhou, Yangqiaoyu  and
      Liu, Haokun  and
      Srivastava, Tejes  and
      Mei, Hongyuan  and
      Tan, Chenhao",
    editor = "Peled-Cohen, Lotem  and
      Calderon, Nitay  and
      Lissak, Shir  and
      Reichart, Roi",
    booktitle = "Proceedings of the 1st Workshop on NLP for Science (NLP4Science)",
    month = nov,
    year = "2024",
    address = "Miami, FL, USA",
    publisher = "Association for Computational Linguistics",
    url = "https://aclanthology.org/2024.nlp4science-1.10/",
    doi = "10.18653/v1/2024.nlp4science-1.10",
    pages = "117--139"
}
